\documentstyle[prb,aps,preprint]{revtex}
\begin{document}
\preprint{DTP 97--XX}
\draft
\title{ DISSIPATIVE DYNAMICS OF SOLITONS IN PLANAR FERROMAGNETS }

\author{Jacek Dziarmaga\thanks{E-mail: {\tt J.P.Dziarmaga@durham.ac.uk}} }
\address{Department of Mathematical Sciences,
         University of Durham, South Road, Durham, DH1 3LE,
         United Kingdom}
\date{March 4, 1997}
\maketitle
\tighten

\begin{abstract}
Dynamics of magnetic bubbles in planar ferromagnets described by the
Landau-Lifshitz equation with dissipation is analyzed. The pure $O(3)$
sigma model has static multisoliton solutions, characterized by a number
of parameters. The parameters describe a finite dimensional manifold. A
small perturbation of energy functional with respect to the sigma model
forces solitons to move. Multisoliton dynamics is effectively reduced to a
flow in the parameter space. 
\end{abstract}

\paragraph{Bogomol'nyi theories.} There exists quite a number of classical
field theories with the so called Bogomol'nyi \cite{bog} energy bound. The
classic example of such a theory, which is familiar to both particle
physics and condensed matter community, is an $O(3)$ sigma model or a
planar ferromagnet \cite{belpol} described in appropriate units by the
energy functional

  \begin{equation}\label{10}
  E_{\sigma}[\vec{M}]=\frac{1}{2}\int d^{2}x\; 
                      \partial_{k}\vec{M}\partial_{k}\vec{M} \;\;,
  \end{equation}
where $k=1,2$ runs over planar dimensions and $\vec{M}$ is a $3$-component
magnetization vector subject to the constraint

  \begin{equation}\label{20}
  \vec{M}\;\vec{M}=1 \;\;.
  \end{equation} 
For the energy (\ref{10}) to be finite the magnetization tends to a
constant vector at spatial infinity for any time. For definiteness we take
$(0,0,1)$ as this constant vector. With this boundary condition
$\vec{M}(t;\vec{x})$ can be viewed as a map from a compactified plane (
equivalent to $S^{2}$ ) to the $S^{2}$ manifold of magnetization defined
by the constraint (\ref{20}).  The energy functional (\ref{10}) is bounded
from below

  \begin{equation}\label{30}
  E_{\sigma}[\vec{M}]\geq 4\pi |N| \;\;.
  \end{equation}
$N$ is an integer topological index of the map $S^{2}\rightarrow S^{2}$,

  \begin{equation}\label{40}
  N=\int d^{2}x\;q(t,\vec{x})\equiv
    \frac{1}{4\pi} \int d^{2}x\; 
    \vec{M}[\partial_{1}\vec{M}\times\partial_{2}\vec{M}] \;\;,
  \end{equation}
where $q(t,\vec{x})$ is a topological charge density. For any winding
number $N$ the bound (\ref{30}) is saturated by a static multisoliton
configuration characterized by a finite number of parameters $\xi$. For
example a solution with a negative topological index of $-n$ is given by

  \begin{eqnarray}\label{60}
  &&M_{B}(\vec{x};c,a_{i},b_{i})=( \frac{W+\bar{W}}{1+|W|^{2}},
                                  i\frac{W-\bar{W}}{1+|W|^{2}},
                                   \frac{1-|W|^{2}}{1+|W|^{2}} ) 
                                                         \;\;,\nonumber\\
  &&W=c\frac{(z-a_{1})...(z-a_{n-1})}{(z-b_{1})...(z-b_{n})} \;\;.
  \end{eqnarray}
$a$'s, $b$'s and $c$ are complex parameters, $z=x+iy$. Not all parameters
are independent, some different combinations of parameters have to be
identified as they give the same $\vec{M}_{B}$. After this identification
there are $4n-1$ real parameters left, they parametrize a
$(4n-1)$-dimensional real manifold $M_{n}$ which will be called a moduli
space. It has to be stressed that (\ref{60}) is a time-independent
multisoliton solution, its energy does not depend on the choice of
parameters.

\paragraph{Relativistic dynamics.} The energy functional (\ref{10}) defines
a static version of the sigma model. Dynamics can be introduced to the
model in a couple of ways. One dynamical version is a relativistic model
described by the Lagrangian

  \begin{equation}\label{70}
  L_{\sigma}[\vec{M}]=\frac{1}{2} \int d^{2}x\;
                   [\;\partial_{t}\vec{M}\partial_{t}\vec{M}-
                      \partial_{k}\vec{M}\partial_{k}\vec{M}\;] \;\;.
  \end{equation}
A method to study low energy soliton dynamics in a relativistic
Bogomol'nyi theory has been proposed by Manton \cite{manton1} and can be
briefly summarized as follows. A Bogomol'nyi solution like (\ref{60})
saturates the lower energy bound (\ref{30}). If solitons are forced by
initial conditions to move with low velocity, a field configuration at any
instant of time remains close to the Bogomol'nyi solution and can be
approximated by $\vec{M}_{B}[\vec{x};\xi(t)]$ with time dependent
parameters.  The approximation is expected to be the better the lower is
velocity. Such an approximate configuration can be substituted to the
Lagrangian (\ref{70}), one obtains after integration over plane an
effective low energy Lagrangian

  \begin{eqnarray}\label{80}
  &&L_{eff}=\frac{1}{2} g_{\alpha\beta}(\xi) 
                      \dot{\xi}^{\alpha} \dot{\xi}^{\beta} \;\;,\nonumber\\
  &&g_{\alpha\beta}(\xi)=
             \int d^{2}x\; 
             \frac{\partial\vec{M}_{B}}{\partial\xi^{\alpha}}
             \frac{\partial\vec{M}_{B}}{\partial\xi^{\beta}} \;\;.
  \end{eqnarray}
The field theory (\ref{70}) is effectively reduced to a finite dimensional
mechanical system. The low energy dynamics of solitons is described by a
geodesic motion on the moduli space $M_{n}$ equipped with the metric tensor
$g_{\alpha\beta}(\xi)$. The geodesic approximation has been studied in
detail for the relativistic sigma model in ~\cite{wardleese}. It has also
been explored in other models like the original BPS theory of monopoles
\cite{manton1} or the abelian Higgs model \cite{samols}.

\paragraph{Landau-Lifshitz equation with dissipation.} Another dynamical
version of the Bogomol'nyi theory (\ref{10}), which of main interest for
us, is given by

  \begin{equation}\label{c10}
  \lambda \;\vec{M} \times \partial_{t}\vec{M}+
  \hat{\Gamma}(\vec{M})\;\partial_{t}\vec{M}=
  \hat{P}_{\vec{M}} [\;\nabla^{2} \vec{M} + 
           \varepsilon \frac{\delta V}{\delta \vec{M}}(\vec{M}) \;] \;\;,
  \end{equation} 
where $\hat{P}_{\vec{M}}$ is a projection operator on a subspace
orthogonal to $\vec{M}$, defined by
$\hat{P}\vec{A}=\vec{A}-\vec{M}(\vec{M}\vec{A})$ for any vector $\vec{A}$.
The equation (\ref{c10}) has to be supplemented by the constraint
(\ref{20}). $\hat{\Gamma}(\vec{M})$ is assumed to be a positively definite
symmetric matrix for generality. $\varepsilon V[\vec{M}]$ is a small
perturbation with respect to the Bogomol'nyi energy (\ref{10}), the total
energy functional is $E=E_{\sigma}+\varepsilon V$. 

  Eq. (\ref{c10}) is satisfied for $\varepsilon=0$ by the static
multisoliton Bogomol'nyi field (\ref{60}),
 
  \begin{equation}\label{c15}
  \nabla^{2}\vec{M}_{B}
  -\vec{M}_{B}(\vec{M}_{B}\nabla^{2}\vec{M}_{B})=0 \;\;.
  \end{equation}
For nonzero $\varepsilon$ the solitons are no longer static, they move
with velocities proportional to $\varepsilon$

  \begin{equation}\label{c20}
  \dot{\xi}^{\alpha}=O(\varepsilon) \;\;,
  \end{equation}
where the index $\alpha$ numbers the collective coordinates.  At the same
time magnetization is given by the Bogomol'nyi field plus a small
deviation

  \begin{equation}\label{c30}
  \vec{M}(t,\vec{x})=
  \vec{M}_{B}[\vec{x},\xi(t)]+ 
  \varepsilon\;\vec{m}(t,\vec{x})+O(\varepsilon^{2}) \;\;.
  \end{equation}

  The equation (\ref{c10}) does not have Lagrangian formulation in a
generic case of $\hat{\Gamma}\neq 0$, one can not proceed along the same
lines as in the relativistic case. Instead one has to rely on field
equations.  Eqs.(\ref{c20},\ref{c30}) define our perturbative expansion in
$\varepsilon$. It follows from the condition (\ref{c20}) that
$\partial_{t}\vec{m}(t,\vec{x})=O(\varepsilon)$. Substitution of
Eq.(\ref{c30}) to the field equation (\ref{c10}) and linearization in
$\varepsilon$ gives

  \begin{eqnarray}\label{c40}
  &&\dot{\xi}^{\alpha}\;[\;
  \lambda\;\vec{M}_{B} \times 
  \frac{\partial\vec{M}_{B}}{\partial\xi^{\alpha}} +
  \hat{\Gamma}(\vec{M}_{B})\;
       \frac{\partial\vec{M}_{B}}{\partial\xi^{\alpha}} \;]+
  \varepsilon\;\hat{P}_{\vec{M}_{B}}\;\frac{\delta V}{\delta\vec{M}}(\vec{M}_{B})=
  \nonumber\\
  &&\varepsilon\;[\;\nabla^{2}\vec{m}
                 -\vec{M}_{B}(\vec{M}_{B}\nabla^{2}\vec{m})-
                  (\vec{M}_{B}\nabla^{2}\vec{M}_{B})\vec{m} -
                  \vec{M}_{B}(\vec{m}\nabla^{2}\vec{M}_{B})\;] \;\;.
  \end{eqnarray}
Similar linearization of the constraint (\ref{20}) leads to a constraint
on $\vec{m}$

  \begin{equation}\label{c50}
  \vec{M}_{B}\;\vec{m}=0 \;\;.
  \end{equation}
$\nabla^{2}\vec{M}_{B}$ is parallel to $\vec{M}_{B}$ according to
Eq.(\ref{c15}). Because of this property and the constraint (\ref{c50})
the last term on the RHS of Eq.(\ref{c40}) is zero. 

  Eq.(\ref{c40}) is a linear inhomogeneous equation for $\vec{m}$. The
source term on the LHS of this equation depends on the Bogomol'nyi fields
only, the RHS can be interpreted as a linear operator (dependent on
$\vec{M}_{B}[\vec{x},\xi(t)]$) acting on the field $\vec{m}(t,\vec{x})$,
say, $\hat{L}\vec{m}$.  A projection of Eq.(\ref{c40}) on
$\frac{\partial\vec{M}_{B}}{\partial\xi^{\beta}}$ somewhat similar as in
~\cite{dorsey}, which is a left zero mode of $\hat{L}$, results in a
solvability condition

  \begin{eqnarray}\label{c60}
  &&\dot{\xi}^{\alpha}\;
    [\;\lambda\;\omega_{\alpha\beta}(\xi)+
     G_{\alpha\beta}(\xi)\;]
    -\varepsilon\; F_{\beta}(\xi)=  \nonumber \\
  &&\varepsilon \int d^{2}x\; 
    \frac{\partial\vec{M}_{B}}{\partial\xi^{\beta}}\;
    [\;\nabla^{2}\vec{m}-\vec{M}_{B}(\vec{M}_{B}\nabla^{2}\vec{m}) -
                (\vec{M}_{B}\nabla^{2}\vec{M}_{B})\vec{m}\;] \;\;,
  \end{eqnarray}
where the tensors on the LHS are defined by

  \begin{eqnarray}\label{c70}
  &&\omega_{\alpha\beta}(\xi)=
  \int d^{2}x\; 
  \vec{M}_{B}( \frac{\partial\vec{M}_{B}}{\partial\xi^{\alpha}} \times
               \frac{\partial\vec{M}_{B}}{\partial\xi^{\beta}} ) \;\;,
  \nonumber \\
  &&G_{\alpha\beta}(\xi)=
  \int d^{2}x\;
  \frac{\partial\vec{M}_{B}}{\partial\xi^{\alpha}} \;
  \hat{\Gamma}(\vec{M}_{B}) \;
  \frac{\partial\vec{M}_{B}}{\partial\xi^{\beta}} \;\;,
  \nonumber \\
  &&F_{\beta}(\xi)= 
     -\int d^{2}x\;   
     \frac{\partial\vec{M}_{B}}{\partial\xi^{\beta}}
     \frac{\delta V}{\delta \vec{M}}(\vec{M}_{B})=
     -\frac{\partial}{\partial\xi^{\beta}}
      \int d^{2}x\; 
      V[\vec{M}_{B}(\vec{x},\xi)] \;\;.
  \end{eqnarray} 
If $\hat{\Gamma}=\gamma\hat{1}$ with a constant $\gamma$, then
$G_{\alpha\beta}(\xi)=\gamma g_{\alpha\beta}(\xi)$ is proportional to the
metric tensor (\ref{80}) on the moduli space, as it was discussed
qualitatively in ~\cite{manton2}. $F_{\beta}(\xi)$ can be interpreted as a
potential force. 

  The RHS of Eq.(\ref{c60}) is zero. Clearly the second term of the
integrand is zero because $\vec{M}_{B}
\frac{\partial\vec{M}_{B}}{\partial\xi^{\beta}}=0$ thanks to the
constraint (\ref{20}). After integration by parts the RHS of
Eq.(\ref{c60}) becomes

  \begin{equation}\label{c100}
  \varepsilon \int d^{2}x\; \vec{m}\;[\;
   \nabla^{2} \frac{\partial\vec{M}_{B}}{\partial\xi^{\beta}}
   - \frac{\partial\vec{M}_{B}}{\partial\xi^{\beta}}
     (\vec{M}_{B}\nabla^{2}\vec{M}_{B}) \;] \;\;.
  \end{equation}
On the other hand taking a derivative
$\frac{\partial}{\partial\xi^{\beta}}$ of Eq.(\ref{c15}) gives

  \begin{equation}\label{c110}
   \nabla^{2} \frac{\partial\vec{M}_{B}}{\partial\xi^{\beta}}
   - \frac{\partial\vec{M}_{B}}{\partial\xi^{\beta}}
     (\vec{M}_{B}\nabla^{2}\vec{M}_{B})
   - \vec{M}_{B} 
     \frac{\partial}{\partial\xi^{\beta}}
      (\vec{M}_{B}\nabla^{2}\vec{M}_{B})=0 \;\;.
  \end{equation}
This equation and the constraint (\ref{c50}) imply that the integral
(\ref{c100}) is zero. 

  To summarize, we have found that a solvability condition for
Eqs.(\ref{c40},\ref{c50}) is given by the following equation of motion for
the collective coordinates

  \begin{equation}\label{c120}
  \dot{\xi}^{\alpha} \;
  [\;\lambda \; \omega_{\alpha\beta}(\xi)+G_{\alpha\beta}(\xi)\;]
  = \varepsilon \; F_{\beta}(\xi) \;\;.
  \end{equation}
Once again the dynamics of a field theory is reduced to a finite
dimensional mechanical system.

  \paragraph{Example.} To substantiate the general discussion by a simple
example let us consider the case of one soliton in an external potential
and $\hat{\Gamma}=\gamma\hat{1}$ with a constant $\gamma$. A general form
of one soliton solution is given by $W=\mu/(z-\nu)$, where $\nu$ is a
complex position of the soliton and the real $\mu$ is soliton's size. 
Nonvanishing tensor elements are

  \begin{eqnarray}\label{c130}
  &&\omega_{\nu\bar{\nu}}=-\omega_{\bar{\nu}\nu}=-2\pi i \;\;,\nonumber \\
  &&g_{\nu\bar{\nu}}=g_{\bar{\nu}\nu}=2\pi \;\;.
  \end{eqnarray}
$g_{\mu\mu}$ is divergent on an infinite plane; it follows from the
$\beta=\mu$ component of Eq.(\ref{c120}) that $\mu$ is constant provided
that $F_{\mu}$ is finite. Let the interaction energy with an external
potential be given by

  \begin{equation}\label{c140}
  V=e\int d^{2}x\; \delta^{(2)}(\vec{x}) \; q(\vec{x},t) \;\;,
  \end{equation}
where $q$ is the topological charge density, which is negative in this
case. For $e>0$ the soliton should be attracted by the impurity at the
origin. This form of interaction energy appears for example in a sigma
model for quantum Hall ferromagnet \cite{qhe}. The potential forces are
 
  \begin{eqnarray}\label{c150}
  &&F_{\mu}=\frac{ 2e\mu (\nu\bar{\nu}-\mu^{2}) }
                 { \pi (\nu\bar{\nu}+\mu^{2})^{3} } \;\;, \nonumber \\
  &&F_{\nu}=\bar{F_{\bar{\nu}}}=
        -\frac{ 2e\mu^{2} \bar{\nu} }
         { \pi (\nu\bar{\nu}+\mu^{2})^{3} } \;\;.
  \end{eqnarray}
The equation of motion for $\nu$ is

  \begin{equation}\label{c160}
  (\gamma-i\lambda)\dot{\nu}=
  -(\frac{e}{\pi^{2}}) \frac{\mu^{2}\nu}{(\nu\bar{\nu}+\mu^{2})^{3}} \;\;.
  \end{equation}
For $\gamma\neq 0$ the solution of this equation is given by

  \begin{eqnarray}\label{c170}
  &&\Theta(t)=\Theta(0)-(\tan B)\ln\frac{R(t)}{R(0)} \;\;, \nonumber \\
  &&2\ln\frac{R(t)}{R(0)}+
    3[R^{2}(t)-R^{2}(0)]+
    \frac{3}{2}[R^{2}(t)-R^{2}(0)]^{2}+
    \frac{1}{3}[R^{2}(t)-R^{2}(0)]^{3}=
    -\frac{2\cos B}{A}t \;\;,
  \end{eqnarray}
where $Ae^{iB}=\pi^{2}\mu^{4}(\gamma-i\lambda)/e$, $Re^{i\Theta}=\nu/\mu$
and $\mu$ is constant. For $\lambda=0$ (purely dissipative case) $\nu$
relaxes to the equilibrium position at $\nu=0$ along a radial line. In
general it moves towards $\nu=0$ along spiral lines. In the case of
Landau-Lifshitz equation or $\gamma=0$ the soliton rotates around the
origin along an equipotential circular orbit

  \begin{eqnarray}\label{c180}
  &&R(t)=R(0) \;\;, \nonumber\\
  &&\Theta(t)=\Theta(0)-\frac{t}{A[1+R^{2}(0)]^{3}} \;\;.
  \end{eqnarray}

  \paragraph{Conclusion.} The central result is the equation (\ref{c120}),
which gives a prescription how to deal with dynamics of solitons in
dissipative system close to the Bogomol'nyi limit. The equation can be
easily generalized to other models because its derivation does not depend
much on the special properties of the theory (\ref{c10}). 

  Generalization to a model with more than one order parameter or field is
possible. For such a model it would be natural to expect the relaxation
times $\gamma_{i}$ for different order parameters to be different,
$\gamma_{i}\neq\gamma_{j}$ if $i\neq j$. In such a case the tensor
$G_{\alpha\beta}$ is not proportional to the metric tensor
$g_{\alpha\beta}$ even if $\gamma$'s are constants. 

  One of applications could be the dynamics of vortices in superconductors
at nonzero temperature. A Bogomol'nyi theory in this case is defined by a
Ginzburg-Landau functional for a superconductor at a border between type I
and type II superconductivity. An appropriate small perturbation of the
quartic potential\cite{shah}, corresponding to $\varepsilon V$, drives the
system in the direction of weak type II superconductivity. Somewhat similar
free energy functional as for superconductors describes transition from
smectic A to nematic phase of liquid crystals \cite{degennes}. 

\acknowledgements

I would like to thank Wojtek Zakrzewski and Bernard Piette for helpful
discussions. This research was supported by UK PPARC.

\end{document}